\documentclass[12pt]{iopart}
\usepackage[dvips]{graphicx}
\begin{document}

\title[Size effects in IR-optical properties of ultrathin Pb quantized films]{Size effects in IR-optical properties of ultrathin Pb quantized films.}

\author{ M Str\'o\.zak$^{1}$, V Hnatyuk$^{1}$$^{,}$$^{2}$, M Ja\l{}ochowski$^{1}$}

\address{{$^{1}$Institute of Physics and Nanotechnology Center, M. Curie-Sk\l{}odowska University}, Pl. M. Curie-Sk\l{}odowskiej 1,
PL-20031 Lublin, Poland}
\address{\mbox{$^{2}$European College of Polish and Ukrainian Universities, M. Curie-Sklodowskiej 25},
PL-20029 Lublin, Poland}

\ead{\mbox{\mailto{viktor.hnatyuk@umcs.lublin.pl},
\mailto{ifmjk@tytan.umcs.lublin.pl},}
\mailto{strozak@tytan.umcs.lublin.pl}}
\begin{abstract}
The reflectance difference (RD) as a function of film thickness was
measured during Pb deposition on Si(111)-($6\times6$)Au surface at
105 K. The oblique incident \emph{s-} and \emph{p-}polarized light
with energy range 0.25-0.60 eV was used. The component of the
dielectric function tensor parallel to the surface was determined
from the data of experiments and it oscillation due to quantum size
effects (QSE) with period of 2 ML within the Pb
monolayer-by-monolayer growth was observed.
\end{abstract}
\pacs{ 78.20.Ci, 78.67.-n, 78.68.+m }
%\submitto{\JPCM}
%\maketitle
\section{Introduction}
It is well-known that optical and electrical properties of thin
films behaive in an unusual way in contrast to physical properties
of  bulk materials. When the thickness of the smooth thin film
become comparable to the de Broglie wavelenght of the electrons,
more subtle effects, such as QSE can be detectible. A number of
experimental works on low-energy electron reflection and
transmission~\cite{art1},\cite{art4},\cite{art5},\cite{art6},\cite{art7},
electron tunneling \cite{art2},\cite{art3}, electrical
conductivity~\cite{art8},\cite{art9},\cite{art10}, valence band
photoemission~\cite{art11},\cite{art12}, electron beam
intensity~\cite{art13}, critical
temperature~\cite{art14},\cite{art15}, Hall
coefficient~\cite{art16},\cite{art17}, work function~\cite{art18}
measurements confirm earlier theoretical
predictions~\cite{art19},\cite{art20},\cite{art21},\cite{art22}.
Influence of  the QSE on growth morphology of Pb/Si(111) systems  is
the subject of the last
investigation~\cite{art23},\cite{art24},\cite{art25}. However,
experimental evidence of the QSE in the optical properties of thin
metallic films still is not so
clear~\cite{art26},\cite{art27},\cite{art28},\cite{art29},\cite{art30}.

Theoretically, the influence of discrete quantum levels on the
optical properties of the cubic particles has been analyzed by Wood
and Ashcroft~\cite{art31}. The formalism based on general random
phase approximation (RPA) applied to simple particle-in-a-box model
and a formula for dielectric constant of small metallic particle has
been derived. The identity of the dielectric function of cubic
particles and that of the film had been noted by Cini and
Ascarelli~\cite{art32}. In this paper, we study optical response
changes during epitaxial growth of ultrathin Pb films on
Si(111)-(6$\times$6)Au substrate by means of  reflectance
differential spectroscopy (RDS).

We used s- and p-polarized light with the incident photon energy
\mbox{range 0.25 - 0.6 eV}. Clear evidence of  QSE peaks with the
period 2 ML of Pb film especially for small photon energies was
observed. The experimental results were analyzed in the frame of the
theory Dignam, Moskovitz and Stobie~\cite{art33}.
\section{Experimental}
The sample preparation and optical measurements were held in UHV
chamber equipped with reflection high energy electron diffraction
(RHEED). A gas-flow UHV liquid nitrogen cryostat and crystal quartz
monitor were used for sample cooling and film thickness measurements
during Pb films deposition process. The based pressure was less than
1$\times10^{-10}$ mbar.

A few direct current flashing  was used to clear substrate and to
produce \mbox{Si(111)-(7$\times$7)} superstructure. In order to
produce Si(111)-(6$\times$6)Au reconstruction, about 1.2 ML of Au
were deposited onto Si(111)-(7$\times$7) and were annealed for 1 min
at about 950 K with gradually lowering temperature to the initial
state.

Optical system was consisted of globar, prism monochromator,
polarizer and PbSe detector. Linearly polarized light was incident
on a sample at oblique angle about 49$^\circ$. A lock-in technique
was used to recover reflected signal. Differential reflection
spectroscopy based on measuring of a relative change in the sample
reflectivity upon thin film deposition:
$\Delta$$R/R=(R^{Pb+Si}-R^{Si})/R^{Si}$, where $R^{Si}$ and
$R^{Pb+Si}$ are reflectivity of a bare substrate and a substrate
covered with Pb atoms, respectively. The stability of the reflected
signal $\Delta$$R/R$ during each measurements was better than
$10^{-2}$.

For better understanding of the growth mode of Pb and for
controlling of the substrate quality RHEED technique had been used.
RHEED oscillations were observed simultaneously with optical
reflectivity experiments. The temperature of the sample during all
time of Pb deposition was maintained at 105 K.
\section {Results and discussion}
\subsection{Reflectance difference data measurements}
\begin{figure}[h]
\begin{center}
\resizebox{1.0\linewidth}{!}{
\includegraphics{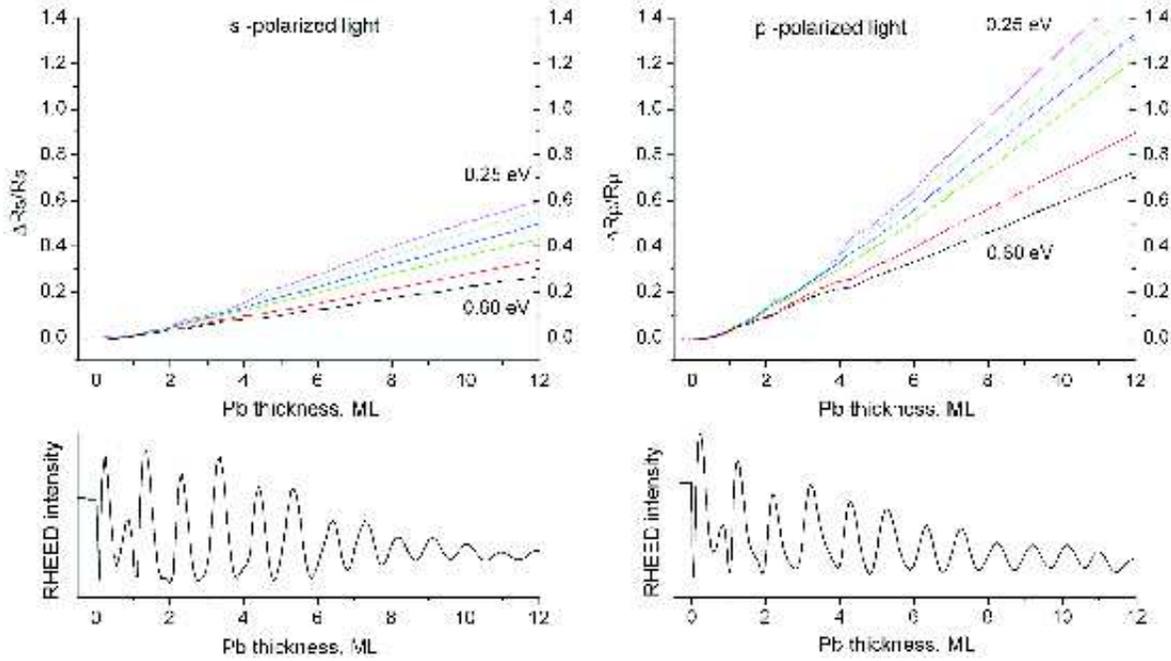}}
\caption{\label{Fig1}DR/R vs Pb film thickness for s-polarized light
(left) and p-polarized light (right) with energies of incident light
h$\nu$=0.25, 0.30, 0.35, 0.40, 0.50, 0.60 eV. Each curve was
measured for a separately prepared thin film sample.}
\end{center}
\end{figure}
Figure 1 presents RD data measured for s- i p-polarized light during
Pb thin film deposition at 105 K onto a Si(111)-(6$\times$6)Au
substrate. The RD curve for energy $h\nu=0.40$ eV increases with the
thickness mostly linearly, while for the rest energies some unusual
features are observed. This deviations from the linearity are more
significant for smaller energies of the incident light. Experimental
data also show different surface response depending on the type of
polarization of the incident light. On the lower panel the RHEED
beam intensity oscillations are presented, which indicates that
films grew in the monolayer-by-monolayer mode~\cite{art13}. The
maximum of the RHEED intensity corresponds the situation when the
layer is complete and surface is atomically smooth.
\subsection{Imaginary part of the dielectric constant and quantum size effect }
A few  theoretical models have been proposed to describe changes in
reflectance of the sample due to thin surface layer formation on the
clear substrate. According to the classical McIntyre and Aspnes~(MA)
model~\cite{art34}, thin film as the homogeneous surface layer can
be described by local and isotropic dielectric function. However,
Feibelman have shown, that this model  could not explain difference
between s- and p-polarization spectrum of the free electron metal,
due to the fact that the z -- component of the electric field in the
surface region has a strong frequency-dependence~\cite{art35}. In
the case of s-polarized light, electric field is parallel to the
surface, varies slowly with going across it and no surface charge is
induced. For p-polarization, the electric field component is normal
to the surface, changes rapidly as one goes across the surface and
can induce a surface charge which in turn can modify surface
response. Since electric fields that are parallel and normal to the
surface responds in the different way, it is necessary to describe
the optical response introducing anisotropic dielectric tensor.

For this purposes we used a classical anisotropic layer Dignam,
Moskovitz and Stobie (DMS) model~\cite{art33}, which assume uniform
anisotropic surface layer to represent thin film on top of the
isotropic substrate. The surface layer can be described by a complex
local dielectric functions $\varepsilon_{z}$ and $\varepsilon_{x}$
as a components normal to and tangential to the surface plane.
Rewriting equation for the relative reflectance changes
$\Delta$$R/R$ of the DMS model in a more convenient way for the case
of Pb thin film on Si substrate, we get to the first order in terms
$d/\lambda$:

\begin{equation}
\fl{~~~~\frac{\Delta
R_{\upsilon}}{R_{\upsilon}}=\frac{8\pi\textrm{}d}{\lambda}\cos\theta
\textrm{~Im}\bigg\{\bigg[\frac{\varepsilon^{Pb}_{x}-\varepsilon^{Si}}{1-\varepsilon^{Si}}\bigg]
\bigg[1+\delta_{\upsilon,p}
\frac{(1-(1/\varepsilon^{Pb}_{z})(\varepsilon^{Pb}_{z}-\varepsilon^{Si})/(\varepsilon^{Pb}_{x}
-\varepsilon^{Si})}{(\cot^{2}\theta-1/\varepsilon^{Si})}\bigg]\bigg\}}\label{1:a}
\end{equation}
where $\upsilon$ = p or s according to the type of polarization of
the wave being considered, $\theta$~is the angle of incidence of the
light beam, $\varepsilon^{Si}$ - the complex bulk dielectric
function of the Si substrate, $\delta_{\upsilon,p}$ is the Kroneker
delta ( equals 1 for $\upsilon=p$, 0 for $\upsilon\neq p$ ). On
setting
$\varepsilon^{Pb}_{z}=\varepsilon^{Pb}_{x}=\varepsilon^{film}$
equation (\ref{1:a}) reduces to equations (24a) and (24b) of MA
model.

Using equation (\ref{1:a}) we calculate imaginary part of the
dielectric function from experimental data presented on figure 1
both for s- and p-polarized light. For Si the bulk dielectric
function after extrapolation to~0.25~eV is equal
$\varepsilon^{Si}$=13+\emph{i}0.001~\cite{art36}. We took
Re$\{\varepsilon^{Pb}_{z}\}$=$Re\{\varepsilon^{Pb}_{x}\}$=~--526~\cite{art37},
but because of imaginary part of the substrate is close to zero,
equation \eref{1:a} is almost insensitive to the real part of the
dielectric function of Pb.
\begin{figure}[h]
\begin{center}
\resizebox{1.0\linewidth}{!}{
\includegraphics{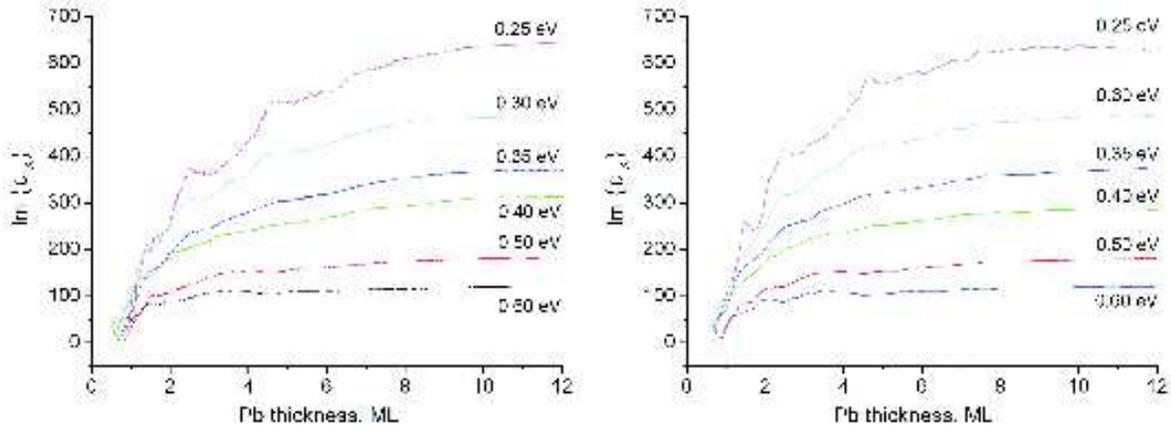}}
\caption{\label{Fig2}Imaginary part of the dielectric function
parallel to the surface Im$\{\varepsilon_{x}\}$ calculated from
experimental data of Fig.1 using equations \eref{1:a} both for s-
and p-polarization.}
\end{center}
\end{figure}

The results for Im$\{\varepsilon^{Pb}_{x}\}$ are shown in figure 2.
The curve with energy 0.40 eV increase with minimal characteristic
changes, but separates the other curves in to two parts. Several
prominent features are observed: for the curves with small energies
of the incident light $h\nu=0.25-0.35$ eV a series of peaks with
periodicity of 2~ML are clearly seen, for the curves with energies
$h\nu=0.50-0.60$ eV an equal number of dips with the same 2~ML
periodicity and positions are also visible. These periodic
oscillations of the Im$\{\varepsilon^{Pb}_{x}\}$ could be
interpreted as the manifestation of the quantum size effect and are
similar to the previously measured surface resistivity
oscillations~\cite{art10}. Since the positions of the peaks and dips
are independent of the energy of incident light, we suppose that
these variations are connected with free electron excitation and are
caused by thickness dependent quantized conducting electron
scattering. All curves saturates at 12 ML, with
Im$\{\varepsilon^{Pb}_{x}\}$ that is typical to the bulk material. A
peak near 1 ML originates from the changes of the surface roughness
induced by the quasi-monolayer by monolayer growth.

According to the quantum theory of Wood and Ashcroft~\cite{art31},
in which discrete energy spectrum of the electron has been taken
into account, the imaginary part of the dielectric function of the
thin metallic film with thickness \emph{d} can be expressed as
follows:
\begin{equation}
\fl{\textrm{Im}\{\varepsilon(x)\}=\bigg(\frac{4}{\pi}\bigg)^4\bigg(\frac{d}{a_{0}}\bigg)\frac{\Gamma}{x}\sum_{n=1}^{n_{c}}
n^{2}(n_{c}^{2}-n^{2})\sum_{{n'=1},{n'\neq
n}}^{\infty}\frac{n'~^{2}[\Delta^{2}+(x^{2}+\Gamma^{2})][1-(-1)^{n+n'}]}{\Delta^{3}[(\Delta^{2}-x^{2}+\Gamma^{2})^{2}+4x^{2}\Gamma^{2}]}}\label{2}
\end{equation}
where $a_{0}$ is the Bohr radius,
\begin{displaymath}
\fl{x=2\hbar\omega\textrm{}m_{eff}d^{2}/(\textrm{}\hbar^{2}\pi^{2}),
\Gamma=(\hbar\textrm{}n_{c}^{2}/E_{F}\tau), n_{c}=Int~(k_{F}d/\pi),
\Delta=n'^{2}-n^{2}}
\end{displaymath}
$E_{F}$ -- is the Fermi energy, $k_{F}$ -- the Fermi wave number,
$\tau$ -- is the relaxation time ($\tau=l/v_{F}$), $l$ -- is the
mean free path, $v_{F}$ -- is the velocity of the electrons on the
Fermi level, $m_{eff}$ -- is the effective mass of the electron and
the function Int takes the integer part of the argument. We
calculate imaginary part of the dielectric function , using next
experimentally obtained parameters of the Pb ultrathin films:
$k_{F}$~=~1.6062 ${\AA}^{-1}$, $l~$~=~20~${\AA}$ and
$m_{eff}$~=~1.002$m_{e}$~\cite{art12}. The results of calculations
\begin{figure}[h]
\begin{center}
\resizebox{0.6\linewidth}{!}{
\includegraphics{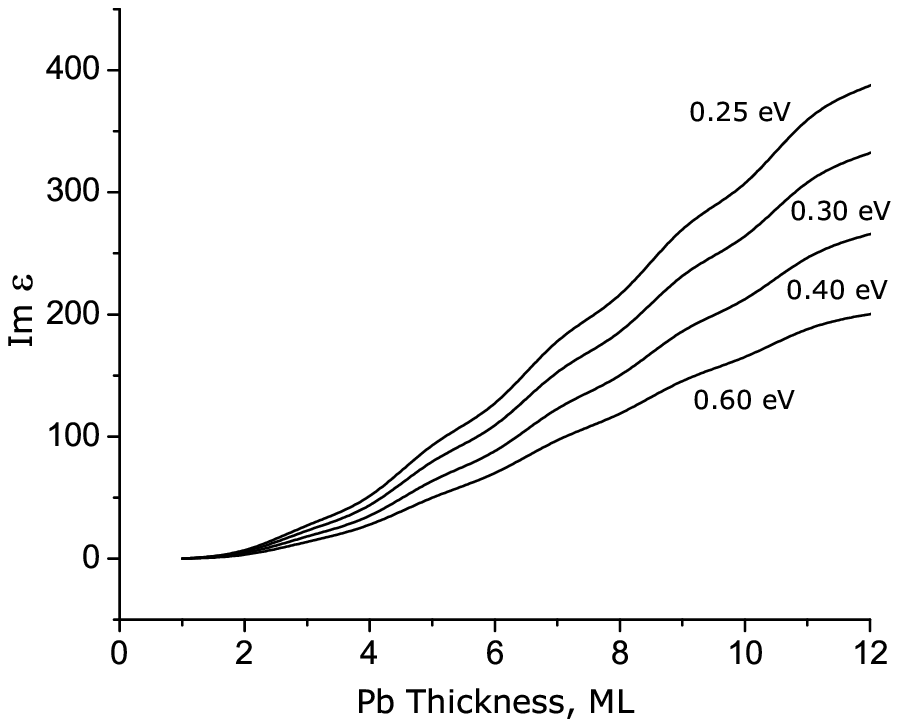}}
\caption{\label{Fig3}Imaginary part of the effective dielectric
function calculated according \mbox{to equation (2) for $k_{F}$ =
1.6062${\AA}^{-1}$,$~l$~=~20~${\AA}$ and $m_{eff}$~=~1.002$m_{e}$.}}
\end{center}
\end{figure}
are shown in figure 3. As it clearly seen Im$\{\varepsilon\}$ growth
vs thickness with periodic oscillations equal to 2 ML. Period of
oscillation is independent of the incident light energy and agree
with ones observed in experimental data. It is satisfy the QSE
condition, when the film thickness \emph{d} is equal to multiple of
one-half the Fermi wavelength  $\lambda_{F}/2$ \cite{art20}, i.e.
$Md_{0}=N\lambda_{F}/2$, where M and N are integers. For Pb(111)
this fulfilled with $(M,N)=(2,3)$~\cite{art9}.
\subsection{Contribution of the electric field parallel and normal to the surface}
Generally it is possible to extract Im$\{\varepsilon^{Pb}_{z}\}$
knowing Im$\{\varepsilon^{Pb}_{x}\}$ from measurements with
s-polarized light, however in our experiments $\Delta$$R_{s}/R_{s}$
and $\Delta$$R_{p}/R_{p}$ were not measured simultaneously and the
sample was changed during the time between experiments, so it is
make problematic to find Im$\{\varepsilon^{Pb}_{z}\}$ with adequate
accuracy. In this case we take the bulk value of the dielectric
constant for the energy $h\nu=0.25$ eV equals
\mbox{Im$\{\varepsilon^{Pb}_{z}\}=$
Im$\{\varepsilon^{Pb}_{x}\}=645$} (figure 2), and define separately
a contribution of Im$\{\varepsilon^{Pb}_{x}\}$ and
Im$\{\varepsilon^{Pb}_{z}\}$ to differential reflectivity of
p-polarized light rewriting equation~(\ref{1:a}) introducing
$(\Delta R_{p}/R_{p})_{x}$ and $(\Delta R_{p}/R_{p})_{z}$
as~follows:
\begin{eqnarray}
\fl~~~\frac{\Delta
R_{p}}{R_{p}}=\frac{8\pi\textrm{}d}{\lambda}\cos\theta\bigg\{\textrm{Im}\bigg[\frac{(\varepsilon^{Pb}_{x}-
\varepsilon^{Si}_{b})(\cot^{2}\theta-1/\varepsilon^{Si}_{b})+(\varepsilon^{Pb}_{x}-1)}{(1-\varepsilon^{Si}_{b})
(\cot^{2}\theta-1/\varepsilon^{Si}_{b})}\bigg]+\nonumber \\
~~~~~~~~~~~~~~~+\textrm{Im}\bigg[\frac{\varepsilon^{Si}_{b}(1/\varepsilon^{Pb}_{z}-1)}{(1-\varepsilon^{Si}_{b})
(\cot^{2}\theta-1/\varepsilon^{Si}_{b})}\bigg]\bigg\}=
\bigg(\frac{\Delta R_{p}}{R_{p}}\bigg)_{x}+\bigg(\frac{\Delta
R_{p}}{R_{p}}\bigg)_{z}
\end{eqnarray}
where first term is the contribution of the surface response when
electric field parallel to the surface, while the second one is the
contribution of the electric field normal to the surface.
\begin{figure}[h]
\begin{center}
\resizebox{0.7\linewidth}{!}{
\includegraphics{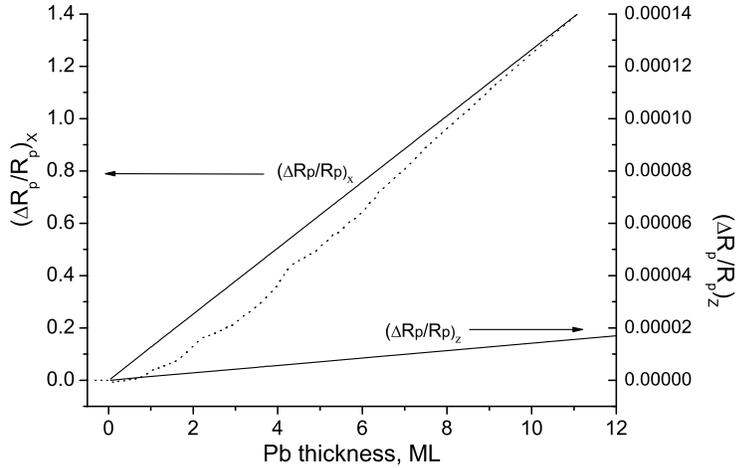}}
\caption{\label{Fig4}Calculated $(\Delta R_{p}/R_{p})_{x}$ and
$(\Delta R_{p}/R_{p})_{z}$ dependences for photon energy $h\nu$=0.25
eV. Dashed curve represents $\Delta R_{p}/R_{p}$  measured during
experiment with the same photon energy (figure 1).}
\end{center}
\end{figure}
Components $(\Delta R_{p}/R_{p})_{x}$ and $(\Delta R_{p}/R_{p})_{z}$
calculated from Im$\{\varepsilon^{Pb}_{x}\}$ and
Im$\{\varepsilon^{Pb}_{z}\}$, respectively for incident photon
energy $h\nu=0.25$ eV are shown in figure 4. Scale of left axis is
$10^{4}$ times larger than the scale of the axis on the left side.
As it clearly seen the main contribution to $\Delta R_{p}/R_{p}$ is
due to the component parallel to the surface $(\Delta
R_{p}/R_{p})_{x}$, while the component perpendicular to the surface
$(\Delta R_{p}/R_{p})_{z}$ is close to zero. This is agree with the
concept of the resistivity measurements, where momentum of
conduction electrons is parallel to the surface of the film. The
behavior of the component normal to the surface is analogous to ones
observed by Borensztein \emph{et al}~\cite{art38} during spectrum
studies of the optical response of clean vicinal Si(001)(2$\times$1)
surface.

It is worth to note, that above mentioned
models~\cite{art33},\cite{art34}, give the same
Im$\{\varepsilon_{x}\}$ dependence with the Pb thickness, calculated
from the data obtained during experiments with the s- and
p-polarized light. We have also analyzed $\Delta R/R$ using
microscopic theory of Bagchi, Barrera and Rajagopal~\cite{art39}.
The results obtained in a long-wavelength approximation are similar
to the results of DMS model for both polarization.

\section{Conclusion}

Optical reflectivity changes of Pb ultrathin films on the
Si(111)-(6$\times$6)Au substrate have been measured using
reflectance difference spectroscopy both for s- and  p-polarized
light. Component of the imaginary part of the dielectric tensor
parallel to the surface has been determined in frame of the
classical Dignam, Moskovitz and Stobie model. It shows strong
thickness dependence up to 10 ML with periodic oscillations equal
2~ML of Pb. These variations in optical properties of Pb thin fims
are due to QSE and are correlated with electrical
resistivity/conductivity oscillations. The contribution of the
component of the imaginary part of the dielectric tensor
perpendicular the surface into surface response is found to be close
to zero in the studied film thickness and photon energy ranges.

\ack One of the author H.V. is grateful to Prof. Miros\l{}aw
Za\l{}u\.zny for helpful discussion.

\end{document}